# Effects of dislocation filtering layers on optical properties of third telecom window emitting InAs/InGaAlAs quantum dots grown on silicon substrates

Wojciech Rudno-Rudziński*[1], Michał Gawełczyk[2], Paweł Podemski[1], Ramasubramanian Balasubramanian[3], Vitalii Sichkovskyi[3], Amnon J. Willinger[4], Gadi Eisenstein[4], Johann P. Reithmaier[3] and Grzegorz Sęk[1]

[1]Department of Experimental Physics, Wrocław University of Science and Technology, St. Wyspiańskiego 27, 50-370 Wrocław, Poland

[2]Institute of Theoretical Physics, Wrocław University of Science and Technology, St. Wyspiańskiego 27, 50-370 Wrocław, Poland

[3]Technological Physics, Institute of Nanostructure Technologies and Analytics, CINSaT, University of Kassel, 34132 Kassel, Germany

[4]Electrical and Computer Engineering Department and Russell Berrie Nanotechnology Institute, Technion-Israel Institute of Technology, Haifa 32000, Israel

**Abstract**

Integrating light emitters based on III-V materials with silicon-based electronics is crucial for further increase in data transfer rates in communication systems since the indirect bandgap of silicon prevents its direct use as a light source. We investigate here InAs/InGaAlAs quantum dot (QD) structures grown directly on 5° off-cut Si substrate and emitting light at 1.5 micrometers, compatible with established telecom platform. Using different dislocation defects filtering layers, exploiting strained superlattices and supplementary QD layers, we mitigate the effects of lattice constant and thermal expansion mismatches between III-V materials and Si during growth. Complementary optical spectroscopy techniques, i.e. photoreflectance and temperature-, time- and polarization-resolved photoluminescence, allow us to determine the optical quality and application potential of the obtained structures by comparing them to a reference sample – state-of-the-art QDs grown on InP. Experimental findings are supported by calculations of excitonic states and optical transitions by combining multiband $\mathbf{k}\cdot\mathbf{p}$ and configuration-interaction methods. We show that our design of structures prevents the generation of a considerable density of defects, as intended. The emission of Si-based structures appears to be much broader than for the reference dots, due to the creation of different QD populations which might be a disadvantage in particular laser applications, however, could be favourable for others, e.g. in broadly tunable devices, sensors, or optical amplifiers. Eventually, we identify the overall most promising combination of defect filtering layers, discuss its advantages and limitations, and prospects for further improvements.

1. **INTRODUCTION**

The applications of light are ubiquitous – from lighting and displays, through sensing, various types of material processing, and, finally, the ever-growing field of data transfer and processing, to name a few. One of the most versatile light sources for these tasks is the laser, especially a very efficient and compact semiconductor one, nowadays usually based on nanostructures, such as quantum wells or quantum dots (QDs). Unfortunately, the most prominent semiconductor compound that dominated the electronic industry, i.e. silicon, is a very poor light emitter on its own because of the indirect band gap. The recent emergence of silicon photonics was made possible only due to the development of technologies that combine efficient light-emitting nanostructures, based predominantly on III-V semiconductors, with silicon substrates. A review of recent advances in self-assembled QD lasers on silicon can be found in Ref. 1.

There are three main methods of integrating III–V materials on Si substrates, including direct growth, bonding, and selective-area hetero-epitaxy.[2] Of these three, the direct growth requires the fewest technological steps, which makes it most economic. However, its main limitation is related to the large lattice constant and thermal expansion mismatches between silicon and practically relevant III-V compounds, such as InAs, GaAs, and InP, which lead to considerable defect formation during epitaxial growth that must be addressed to produce effective light emitters. There were attempts to grow InAs directly in silicon matrix[3], however, QDs obtained in this way showed no light emission. More successful approaches require separation between III-V materials and silicon. It can be achieved by deposition of a very thick buffer layer, which is not very efficient. A better solution, applied in the structures investigated here, is to deposit defect filtering layers (DFL), comprising, e.g., strained superlattices (SLS) or additional QD layers, that prevent the propagation of generated defects to the active region.[4-7] They operate by introducing strain fields that bend the direction of dislocation propagation with Peach–Koehler forces, preventing them from reaching the optically active region.[8]

Most of the efforts so far have been devoted to III–V QD lasers on Si with emission at 1.3 μm, corresponding to the local minimum of optical-fiber losses referred to as the second telecom window. Devices with excellent properties, such as low threshold currents and high operation temperature, were demonstrated.[9-11] However, it is highly desirable to transfer these achievements to the global minimum at 1.55 μm, in the middle of the third telecom window. Therefore, we present here the optical investigations of three InAs/InGaAlAs QD structures designed to achieve efficient emission around 1.5 μm. They are grown on 5° off-cut silicon substrates to avoid creation of antiphase domain boundaries appearing when polar material is grown on a non-polar one. For this to succeed, InAs should be deposited directly on a material lattice-matched to InP, taking advantage of the well-developed growth methodology of InAs/InP QDs.[12-14] The growth of QDs emitting at 1.5 μm is very challenging, due to the large lattice incompatibility between InP and Si of 8% (as compared to 4% for GaAs/Si). Thus, a multiple-step strain relaxation technique is implemented to accommodate lattice constant shift from Si to InP.[15] As DFLs, we use different combinations of SLS or QDs, and we check their influence on the emission properties of the active InAs/InGaAlAs QDs by comprehensive optical characterization, including absorption-like and temperature-dependent emission measurements, as well as studies on emission dynamics and its polarization dependence. As a reference, we study an InAs/InGaAlAs sample grown on an InP substrate to relate the inferred optical properties of Si-based samples to the technologically well-established InP-based platform. Our results reveal the optical quality of QDs grown on Si-substrate. On the one hand, it allows determining the most effective combination of DFLs. On the other, the comparison with the InP-based structure evaluates the application potential of investigated samples as an active material for lasers and indicates the potential for improvements in the growth procedure.

## 2. DESCRIPTION OF SAMPLES

The investigated samples have very complex layouts, therefore only the layer thicknesses and compositions for each structure will be presented, without growth temperatures and deposition rates. More details on the growth procedure can be found in Ref. 16. All the Si-based structures are grown by a Varian Gen II solid-source molecular beam epitaxy (MBE) system on three-inch n-type Silicon (100) wafers, oriented by 5° towards the <110> direction. The general layout of Si-based samples and detailed compositions of each section are shown in Fig. 1. On top of an off-cut Si substrate, two DFLs are deposited, one shifting the lattice constant from Si to GaAs, and the other further from GaAs to InP. The DFLs consist of either superlattices or QDs and the exact composition for each of the three structures is given in Tab. I. An identical section containing optically active InAs QDs is grown on top of the DFLs for all the samples.

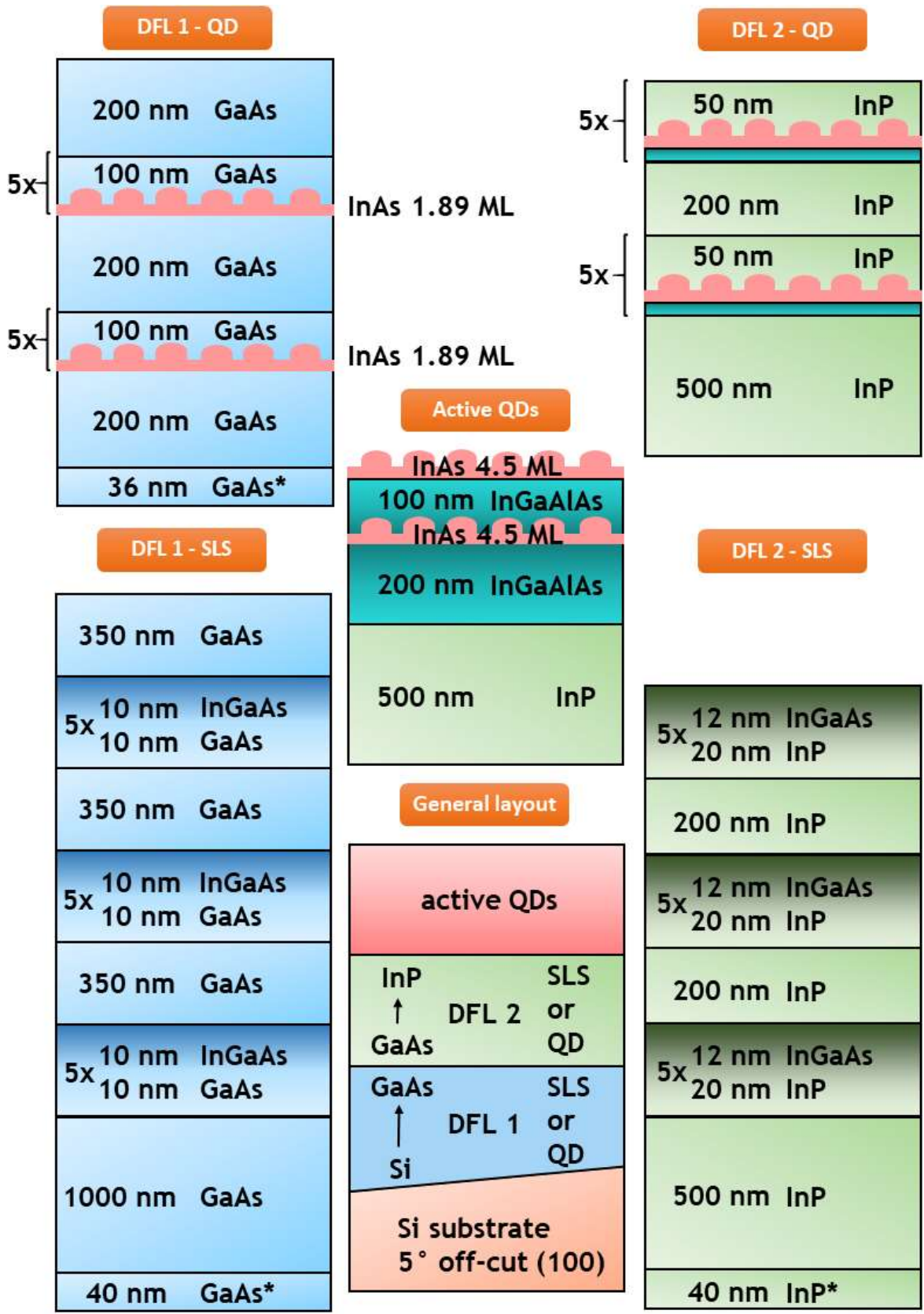

Fig. 1. Schematic representatibon of layer layout of all the investigated samples. For the exact composition of a given sample, please refer to Tab. I

The superlattice GaAs DFL (DFL 1) comprises three sets of 5-period $In_{0.82}Ga_{0.18}As$/GaAs 10/10 nm SLSs, separated by 350 nm thick GaAs layers and deposited on a 1000 nm thick GaAs layer, whose growth is initiated by a thin 40 nm GaAs nucleation layer, deposited at lower temperature and rate. The composition of InP superlattice DFL (DFL 2) is analogous, with three sets of 5-period $In_{0.6}Ga_{0.4}As$/InP 12/20 nm SLSs, separated by 200 nm of InP, with 500 nm thick InP layer at the bottom. The QD GaAs DFL is based on two groups of five layers of InAs QDs, nucleated from 1.89 monolayers (ML) of InAs material, separated by 100 nm of GaAs, with 200 nm of GaAs deposited between the two groups. The InP QD DFL again contains two groups of five repetitions of QDs, but this time the dots are grown on a 2 nm thin InGaAlAs layer lattice matched to InP and covered by 50 nm of InP. The nominal thickness of the InAs QD layer is 5 ML. Two QD groups are separated by 200 nm of InP. Three Si-based samples are investigated, termed A, B, and C.

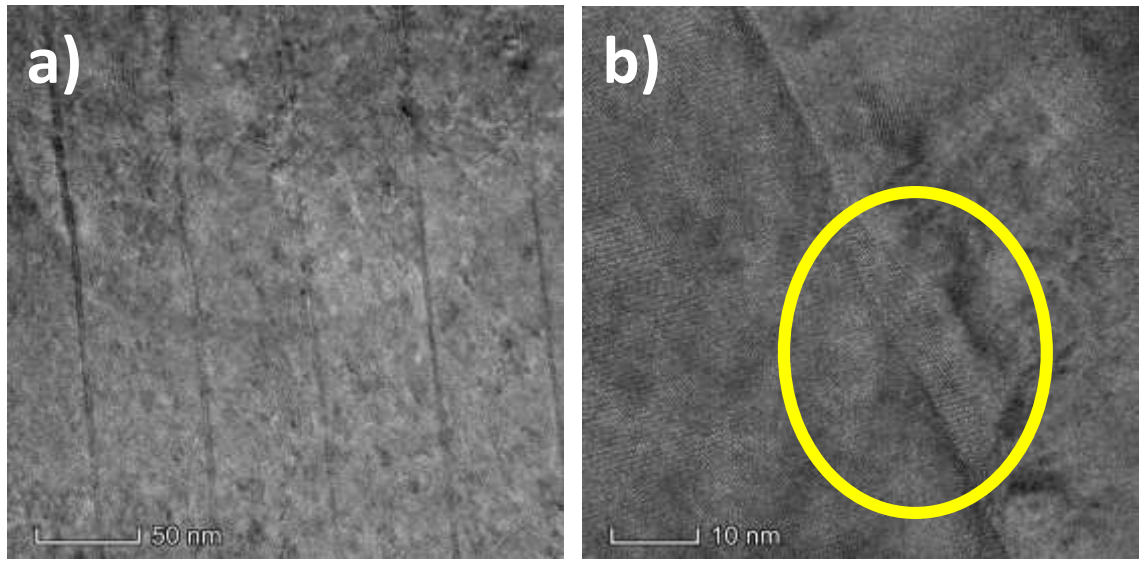

Fig. 2. Cross-sectional TEM images for sample B. a) Upper QD-based DFL b) Zoom on active QD layer, yellow circle indicates one of dots.

The sample A consists of two SLS DFLs, the sample B has GaAs SLS DFL and InP QD DFL (a cross-sectional transmission electron microscopy (TEM) image for the upper part of QD DFL is shown in Fig. 2a)), while the sample C comprises two QD DFLs. On top of the InP DFL layer, a section containing active QDs is grown ("Active QDs" section in Fig. 1). For all the Si-based samples it starts with 500 nm of InP material, on top of which 200 nm $In_{0.53}Ga_{0.23}Al_{0.24}As$ barrier lattice-matched to InP is deposited. The dots nucleate in a 4.5 ML thin InAs layer. Due to the use of the cracked version of arsenic ($As_2$), instead of typical $As_4$, which affects the diffusion on the surface, obtained QDs are almost in-plane symmetric, despite the natural tendency for considerable elongation in the MBE growth of InP systems.[17] A cross-sectional TEM image of an active QD in the sample B is shown in Fig. 2b). QDs are covered with 100 nm of an $In_{0.53}Ga_{0.23}Al_{0.24}As$ barrier, on top of which nominally identical 4.5 ML thin

InAs layer is deposited, leading to the nucleation of unburied QDs carried out to facilitate atomic force microscopy imaging.

Tab. I. Schematic presentation of sample composition.

|  | Sample name | | | |
| --- | --- | --- | --- | --- |
|  | A | B | C | REF |
| DFL 2 | SLS | QD | QD | - |
| DFL 1 | SLS | SLS | QD | - |
| Substrate | Si | Si | Si | InP |

As a reference, there was used a standard QD structure with just an active region as in Fig. 1, grown by MBE under similar conditions on a (100) oriented n-type InP:S substrate. The design of the reference sample is very simple: on top of a thick InP buffer layer, a 100 nm quaternary $In_{0.53}Ga_{0.23}Al_{0.24}As$ layer is deposited, followed by 4.3 ML of InAs material resulting in the formation of QDs and then capped with 100 nm of quaternary $In_{0.53}Ga_{0.23}Al_{0.24}As$ material.

3. **EXPERIMENTAL SETUPS AND COMPUTATIONAL METHODS**

Photoreflectance (PR) spectra were measured in the bright setup configuration, with a whole spectrum of white light, generated by a halogen lamp, reflected off the sample and then focused on a slit of a 23-cm-focal-length monochromator and registered by an InGaAs photodetector connected to a lock-in amplifier. As a source of modulation, the 660 nm line of a semiconductor laser was used, with 10 mW power measured in front of the sample, modulated by a mechanical chopper with the frequency of 273 Hz. More details on the photoreflectance setup can be found elsewhere.[18]

For the PL experiment a similar setup was used, with non-resonant excitation by the 532 nm second-harmonic-emission line of a neodymium-doped yttrium-aluminum-garnet laser, with 10 mW power measured in front of the sample. The shorter excitation wavelength (532 nm, as compared to 660 nm in PR) was applied to decrease the laser penetration depth so that the potential emission from deeper parts of the structure (e.g. QDs in the DFLs) is diminished. The light was focused on the samples with a 10-cm-focal-length lens, leading to excitation power density on the order of 200 W/cm$^2$. The detection was realized in a homodyne scheme, with mechanical modulation at 273 Hz and a signal processed by a digital lock-in amplifier. For temperature series, the samples were mounted on a cold finger in a closed cycle helium refrigerator, providing the temperature range from 13 to 300 K (0.5 K temperature setting accuracy). To realize linear-polarization-sensitive PL experiment, a broadband linear polarizer was introduced in front of the monochromator slit and set to an angle corresponding to its maximum throughput, thus neglecting the polarization characteristic of a diffraction grating. The rotation of an achromatic half-wave plate positioned in front of the polarizer provided the tuning of the linear polarization of detected light.

Time-resolved experiments were realized in a time-correlated single photon counting scheme, with pulsed excitation provided by a Q-switched laser diode, with the emission wavelength of 805 nm, 0.7 mW average power, 40 MHz repetition rate, and around 100 ps pulse length. A small part of the excitation light was split to a trigger diode to facilitate precise time synchronization. An NbN superconducting single photon detector was used to register the arrival of emitted photons, and a time-correlated single photon counter measured the time interval between the synchronization signal and the emitted photons' arrival. Total time resolution of the setup is above 100 ps.

To support our interpretation and conclusions, we theoretically model the QDs under study. For this, we supplement the available information on the morphology from TEM images (as the example in Fig. 2b) with premises from the results of optical measurements, which we will present later, and establish the following model of a QD. We assume a typical QD to have a dome shape with 7.5 nm height based on TEM, and an elliptical base with 20 nm and 40 nm semiaxes, which gives in-plane asymmetry characterized by a lateral aspect ratio of 2 (very often some shape asymmetry is present for this kind of dots.[19] For this QD geometry, settled on a 1.2-nm-thick wetting layer (WL), we initially assume homogeneous QD and WL composition with slight barrier material admixture (~6% of Al and Ga in the QD material), also typically occurring in MBE-grown QDs of this material system.[20] Next, we simulate material diffusion at interfaces by performing Gaussian averaging of the 3D material composition profile with an in-plane spatial extent of 1.8 nm and 0.9 nm in the growth-axis direction. The finally chosen QD morphology is the result of a series of preliminary calculations that allowed us to fine-tune the parameters to match the results of calculations with the spectroscopic observations (e.g. the ground state transition energy, degree of linear polarization).

For this QD, we calculate the strain field by using the theory of continuous elasticity and minimizing the elastic energy of the system on a uniform Cartesian grid of points. The material is non-centrosymmetric, so the shear strain generates a piezoelectric field, which we calculate to the second order in strain. To find the states of electrons and holes in a QD, we use a state-of-the-art implementation[21] of the envelope-function multiband $\mathbf{k}\cdot\mathbf{p}$ theory[22], including the strain, the piezoelectric field, and the spin-orbit interaction. The Hamiltonian used is provided in Ref. 23, and details of the QD modeling and material parameters are given in Ref. 24 and references therein. We numerically diagonalize the Hamiltonian to obtain the single-particle energy levels and carrier eigenstates. To calculate the exciton states, we use the configuration-interaction approach with a configuration basis constructed of 24 electron and 24 hole states. The calculation includes the Coulomb interaction and phenomenological electron-hole exchange interaction. Next, within the dipole approximation[25], we calculate the interband optical transition dipole moments, which gives us information on radiative lifetime and polarization of emission.

**4. RESULTS**

The investigated samples are intended as an active material for lasers, therefore our experiments will be focused on the optical properties of the whole ensemble of dots, mainly their optical quality evidenced by the efficiency of emission; distribution of sizes, and compositions reflected in the broadening of emission and the strength of carrier confinement as well as the efficiency of carrier losses probed by thermal quenching of PL. Additional information will be provided by the analysis of the spectral dispersion of measured luminescence decays and polarization characteristics of emission.

**A. Optical transitions and residual strain**

The growth of self-assembled QDs is predominantly governed by lattice mismatch between dot material and surrounding barrier material. If the two-stage strain relaxation scheme in the Si-based structures works exactly as intended, there should be no difference in strain, and thus structural and optical quality in the active region, between these and the InP-based samples. However, if the strain is not fully relaxed, it should primarily be reflected in the bandgaps of both the InGaAlAs barrier and the thick InP layer. Although the total thickness of Si-based structures is several micrometers, the consecutive layers keep undulating, plus, some of the extended defects can still reach the active region apparently (see Figs. 5 and 6 in Ref. 16) - both will affect the QD nucleation process. Moreover, since they are not uniform over larger areas, they can increase the inhomogeneity of the Si-based QD ensembles.

We use photoreflectance, as an absorption-like modulation technique sensitive to optical transitions even at room temperature (RT),[26] to probe all the major transitions in the entire structure. Although the investigated samples have very complex layouts, the limited penetration depth of the modulating laser considerably reduces the response of the layers below the 500 nm thick InP layer, causing no fingerprints of the DFLs in the optical spectra. The comparison of all the measured RT PR spectra, together with PL results obtained during PR measurements (very low excitation regime – power density below 1 W/cm$^2$) are shown in Fig. 3. To facilitate quantitative analysis, we fit the relevant transitions with the function given in Eq. (1), which is a convenient approximation of Aspnes formula[27] for PR line shape for bulk-like transitions at room temperature:

$$\frac{\Delta R}{R} = \frac{D}{((E-E_i)^2+\Gamma^2)^{3/2}\cos\left(\varphi-3\left(\frac{\pi}{2}-\mathrm{atan}\left(\frac{(E-E_i)}{\Gamma}\right)\right)\right)}, \quad (1)$$

where $E_i$ denotes the transition energy, $D$ is its relative intensity, $\Gamma$ stands for the broadening and $\varphi$ represents the experimental phase, which has no simple physical representation ans is related to the modulation in the lock-in detection scheme.

In the QD region for Si-based samples, no clearly distinctive PR response can be seen, as is very often the case for QD samples, due to the very small total absorption of dots, further diminished by large inhomogeneous broadening.[28] The PR signal for the sample A in the QD emission energy range has an oscillatory character, not related to the QD transition. In the case of the reference sample, a QD-related PR feature could be resolved around 1.5 μm (~ 0.8 eV) as intended (which we relate to better ensemble homogeneity mainly), therefore for this case we also show the respective fitting curve (red line). This PR feature is accompanied by an evident PL peak, indicating that the reference sample has better optical quality at RT apparently.

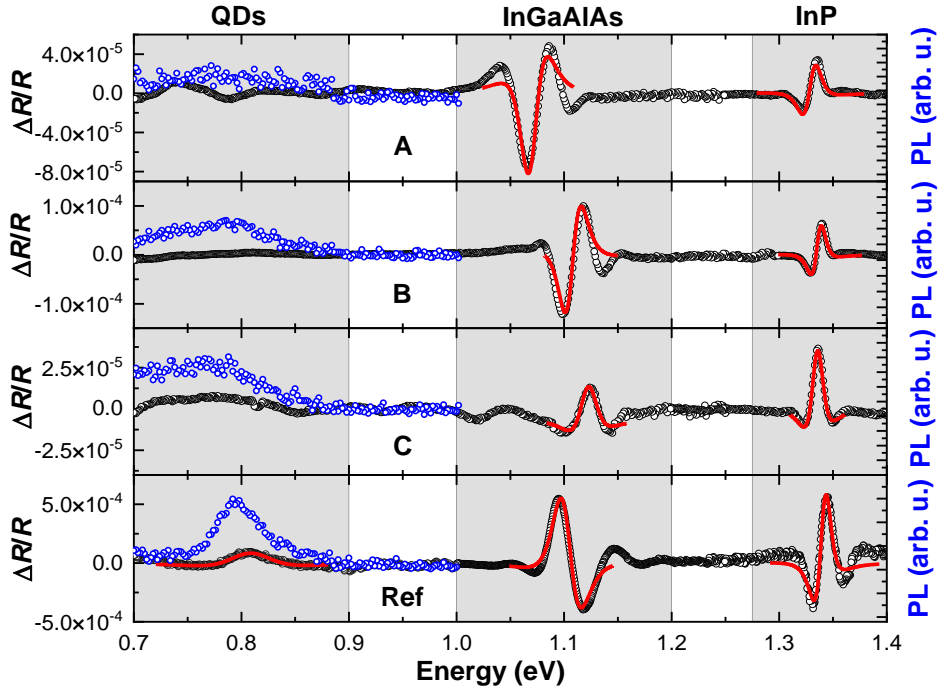

Fig. 3. Room temperature PR spectra (black circles) for all the investigated samples. Open blue circles show PL response. Red lines represent fits to PR spectra according to Eq. (1). Shaded areas indicate spectral regions related to optical transitions in QDs, InGaAlAs barriers, and InP layer.

We summarize the results of the fitting procedure for the bulk-like transitions in Tab. II. The PR resonance attributed to the InP bandgap transition for all the samples is comparable, with a broadening of ~13 meV, which is reasonable for good-quality epitaxial bulk layers measured at RT. There are small differences in its determined energy. For the reference sample, it agrees well with the established value of the InP bandgap of 1.344 eV. It gets only slightly lower for the samples B and C, but reaches 1.329 eV for the sample A. This reduction may be related to some residual strain left in the sample with SLS DFLs. Analyzing the InGaAlAs barrier-related transition is more complex, because of two factors: there are actually two quaternary layers, differing in the growth conditions since the one above the QD layer is affected by its deposition; and minute changes in the composition of InGaAlAs material may affect its band gap. Therefore, the obtained transition energies may not reflect the strain conditions exclusively. For nominal composition, the room temperature bandgap of InGaAlAs is equal to approximately 1.1 eV. Having that in mind, we notice that the barrier energy reveals a significant shift only for the sample A, which agrees with the shift in the InP layer energy for the same sample. Concerning the broadening, whose analysis is free from the abovementioned caveats, Si-based samples show values even slightly smaller than the reference, proving the good quality of the barrier material. Based on those observations, we find the properties of the samples B and C to be closer to the reference.

Tab. II. Results of fitting the function given in Eq. (1) to PR spectra. Uncertainties are related only to the accuracy of the fitting procedure.

|  | **InGaAlAs barrier** | | **InP** | |
|---|---|---|---|---|
|  | **Energy (eV)** | **Broadening (meV)** | **Energy (eV)** | **Broadening (meV)** |
| **A** | 1.070±0.001 | 21.5±0.8 | 1.329±0.001 | 15.6±0.9 |
| **B** | 1.107±0.002 | 17.9±1.2 | 1.336±0.001 | 11.3±0.5 |
| **C** | 1.122±0.001 | 20.1±0.1 | 1.335±0.001 | 13.7±0.6 |
| **Ref** | 1.104±0.001 | 22.5±0.5 | 1.341±0.001 | 13.0±0.9 |

**B. Photoluminescence from quantum dots**

As the most important for structures, designed as an active material for laser application, are their emission properties, in this section we provide an analysis of the PL spectra at room and low temperatures. RT PL spectra are shown in Fig. 4 a). The emission for the reference sample is the most intense (at least its peak intensity), with the broadening of 26 meV, indicating a very homogeneous distribution of dot characteristics and very good ensemble quality for this material system.[29] Its maximum is right at 1.55 µm, in the the telecom C-band, with shoulders reaching also L- and S-bands. In the case of Si-based samples, the total integral intensity is only 20-30 % lower than the reference for the samples B and C, but the intensity at the peak maximum is much lower due to considerably larger broadenings of around 90-100 meV, resulting from increased inhomogeneity of these QD ensembles. The emission from the sample A is the weakest, indicating that the SLS-based DFLs lead to lower quality of QDs. The maxima of emission for all the Si-based samples overlap with telecom bands, however considerable part of the emission reaches longer wavelengths. Such QDs, however, could have practical potential, e.g. for applications requiring broad gain, especially tunable lasers or semiconductor optical amplifiers, yet a very broad emission band forces improvements in dot homogeneity for more typical laser devices, requiring high gain, for instance.

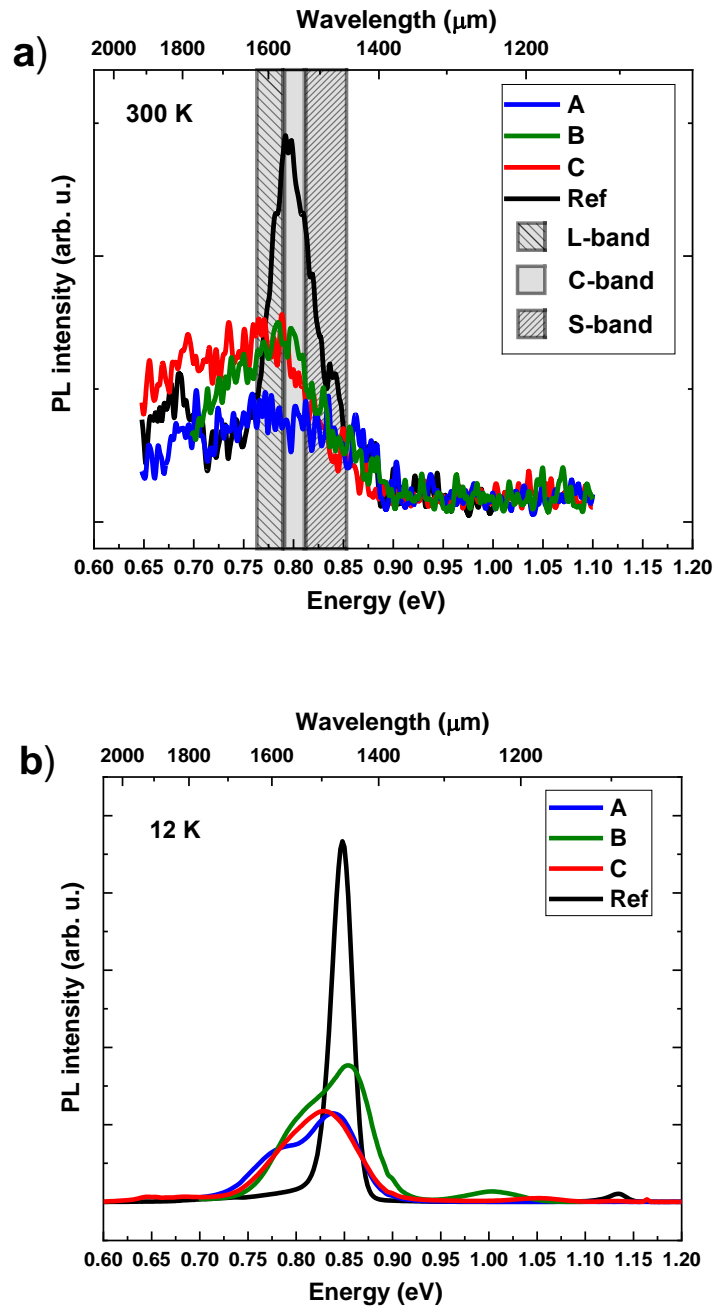

Fig. 4. PL spectra of all the investigated samples measured at a) RT (300 K) and b) low temperature (12 K). The practically relevant region of telecom windows is shaded in a).

To derive more information on the origin of emission peaks, we also measured PL spectra at a low temperature of 12 K, and the results are presented in Fig. 4. b). Besides the obvious temperature-related shift in energy, the overview of the spectra is analogous to the RT results, with changes in relative intensities of Si-based samples to be discussed afterward. What becomes clear is that the emission of the samples A, B, and C comprises at least two peaks, separated by more than 50 meV. The low-energy one (below 0.8 eV at this temperature) evolved from the low-energy emission tail seen already in the room temperature PL spectra. The main, higher energy peak corresponds to the emission in the reference sample. Such behavior of the PL spectra with always coexisting two emission peaks does not correspond to the typical state-filling effect. Therefore, we suppose that it is rather related to different populations of dots due to bimodal size distribution than the ground and excited states of the same QD ensemble. Especially since the peaks' separation does not fit the calculated *s-p* shells splitting of about 35 meV – see the next section. Further indications on the existence of sub-ensembles will be provided later by the spectral dispersion of decay times.

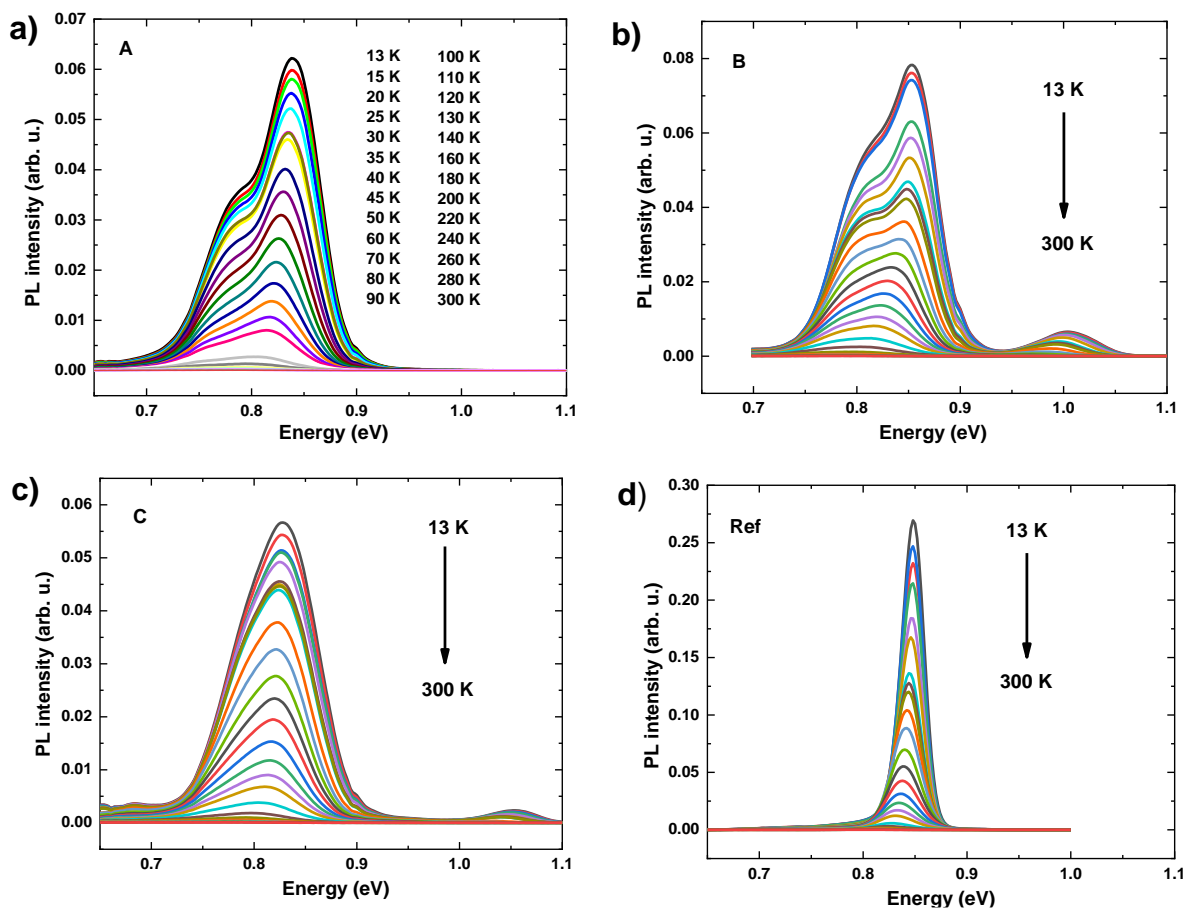

Fig. 5. PL spectra for all the samples measured for different temperatures (listed in a)).

### C. Simulation results

We precede the presentation of the experimental findings with a summary of the results of our modeling given in Tab. III (made for low-temperature material parameters). Referring to them will allow us for a more in-depth interpretation of the results shown below.

Tab. III. Results of numerical simulations of QD states at low temperature.

|  | $X_{1,1}$ (ground state) | $X_{1,3}$ | $X_{1,4}$ | $X_{2,2}$ (*p* shell) |
|---|---|---|---|---|
| Energy (meV) | 841 | 859 | 863 | 876 |
| Relative energy (meV) | 0 | 18 | 22 | 35 |
| Lifetime (ns) | 1.17, 1.80 | ~120 | ~118 | 2.14, 2.50 |
| DOP | 0.21 | 0.24 | −0.69 | 0.08 |

We refer to single-particle states as $e_i$ and $h_i$, where *i* is the state number, and accordingly we call $X_{i,j}$ an exciton state predominantly composed of $e_i$ and $h_j$. The fundamental low-temperature QD optical transition at approximately 0.84 eV corresponds to the $X_{1,1}$ excitonic state. The single-particle splittings between the *s*- and *p*-shell (ground and excited) states are rather low, 23.9 meV for the electron and 8.4 meV for the hole. These values allow us to estimate the distance to the exciton *p*-shell bright states ($X_{2,2}$) to be about 32.3 meV, which is confirmed by direct calculation of the excitonic state spectrum with the *p*-shell state ~35 meV above the ground state. However, given the much smaller hole state splitting, other partly bright states with lower energies are also possible. The lowest of them comprising a ground-state electron and a hole on its second excited state ($X_{1,3}$) is found 18 meV above $X_{1,1}$, and an analogous one involving the third hole excited state is obtained 4 meV further apart. However, both these are characterized by significantly smaller oscillator strength than the $X_{1,1}$ and $X_{2,2}$ states (by approx. two orders of magnitude – see the respective calculated lifetimes in Table III).

The $X_{1,1}$ and $X_{2,2}$ states have two bright spin configurations each, for which we give the radiative lifetimes. The two transitions are almost ideally polarized along [1-10] and [110] directions (often called *V* and *H*) and their unequal lifetimes give rise to some degree of linear polarization for their unresolved emission, here defined as DOP = $(\tau_H - \tau_V)/(\tau_H + \tau_V)$. For $X_{1,3}$ and $X_{1,4}$, we give lifetimes of their brightest spin configurations, and cumulative DOP values for their entire fine-structure multiplets.

**D. Thermal stability of emission**

We use the temperature dependence of PL to estimate carrier confinement strength and reveal potential carrier escape channels. Although structural studies of similar structures[16] estimated low surface density ($10^8$ cm$^{-2}$) of defects in the vicinity of optically active QDs, they still may affect the PL response, especially its temperature stability. Other differences between the samples may be related to the small changes in strain conditions and layer orientations, influencing QD growth and thus their geometric parameters and composition, as mentioned in the previous sections. To shed more light on the strength of confinement potential and non-radiative recombination channels, we measured temperature-dependent PL for all the samples, with the results presented in Fig. 5. The most striking difference between the reference and Si-based samples is the broadening of emission, considerably larger for the ones grown on Si substrates. Due to this fact, although the maximum intensity for the reference sample is much higher than for the rest, their total emission, measured as an integral under the whole area of the PL peaks, is only 2-3 times lower at low temperature. To facilitate reliable comparison of the results between the samples, we integrate the whole QD emission bands for each sample. The obtained temperature dependences are plotted together in Fig. 6.

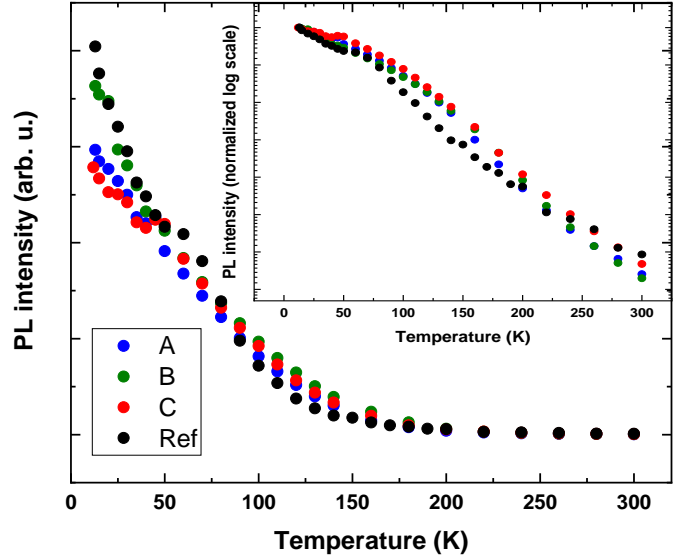

Fig. 6. PL intensity obtained by integrating PL peaks as a function of temperature. The inset shows PL curves in log scale normalized to the lowest temperature values.

As can be seen, the decrease of the total PL intensity for all the structures is similar, and even slightly faster in the case of the reference, especially in the low-temperature range. It can be resolved better in the inset of Fig. 6., showing PL intensities in log scale normalized to the highest low-temperature values. It proves that the optically active QDs in the Si-based samples are of good structural quality, and defect states do not play any significant role in the emission processes. The differences between the samples must be attributed then to the changes in the confinement strength, defined as the energy distance from the ground state to the nearest efficient carrier escape channel (state), which depends on the morphology of dots and details of band structures of the layers surrounding them, including the wetting layer.

We provide a quantitative analysis of temperature dependence of PL intensities by fitting with the relation:

$$\ln I = \ln \frac{I_0}{1+C_1 e^{-\frac{E_1}{kT}}+C_2 e^{-\frac{E_2}{kT}}}, \quad (2)$$

with two activation energies attributed to two different carrier escape routes (a single activation process could not reflect the experimental data). The results are plotted in Fig. 7, and the obtained energies are shown in Tab. IV. It is worth mentioning that the activation energies are obtained for a response averaged over the entire population of emitting QDs. Having that in mind, we may conclude that the lower activation energy for all the samples is comparable and relates to the distance to the lowest bright excited states in a QD ($X_{1,3}$ and $X_{1,4}$ - see Tab. III in Sec. IV. C). According to our calculation, these energies are about 18 and 22 meV, which agrees well with the lower of the experimentally determined activation energy values. The promotion of carriers to these levels reduces the total PL intensity even though the respective emission is still collected in the integrated signal. The reason is that these states are predominantly composed of a ground-state electron and a hole in its second or third excited state, and hence the exciton recombination is much slower than for the ground-state exciton, according to our calculations up to 2 orders of magnitude. The higher activation energy is most probably related to the escape of holes to the quaternary barrier which needs lower energy than for electrons (approx. 80 meV from the calculations for our model QD) as it was also observed for InAs/InGaAlAs quantum dashes on InP.[30] It might slightly differ between the Si-based samples and the reference one, for which it is a bit lower, at least because the carrier confinement in the QD with respect to the barrier band edges can be shared differently between the conduction and valence band when details of the strain distribution, compositions, and morphologies are changed.

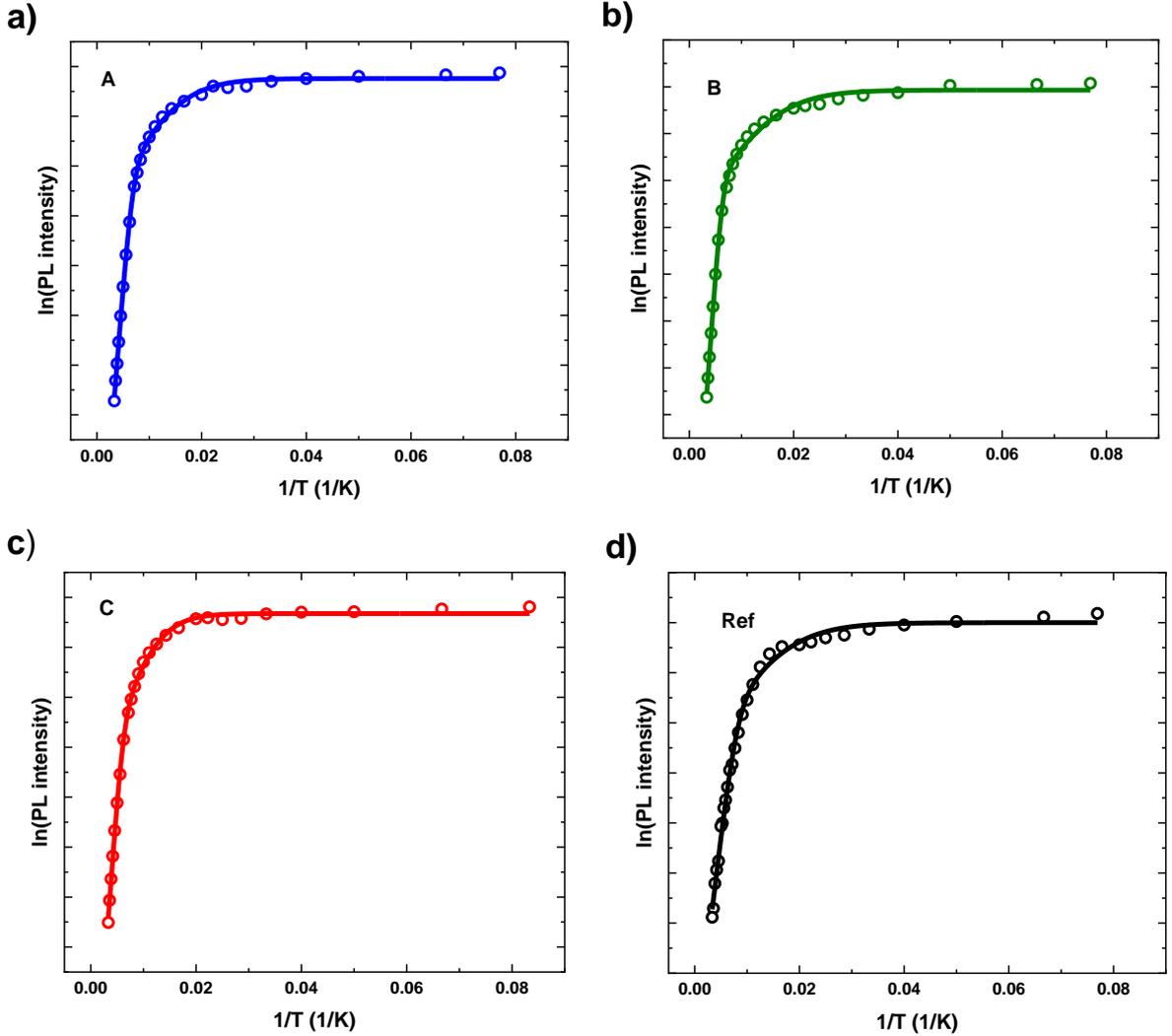

Fig. 7. Arrhenius plots of PL intensities together with fits according to Eq. (2).

Tab. IV. Activation energies determined from fits to Arrhenius plots of PL intensities.

| Sample name | $E_1$ (meV) | $E_2$ (meV) |
|---|---|---|
| A | 18 | 117 |
| B | 16 | 136 |
| C | 26 | 128 |
| Ref | 15 | 78 |

### E. PL dynamics

Low-temperature PL decay times bring important additional information on the emission potential of QDs since they can be directly related to the radiative lifetimes, i.e. a key characteristic of active material translated also into the radiative emission efficiency. The length of a pulse, its average power, and repetition rate in the time-resolved measurements reported here are chosen in such a way that we should generate less than one electron-hole pair per dot during each impulse, meaning that the low-temperature PL is dominated by the excitonic emission (i.e. minimizing the contribution from more complex charge carrier complexes and higher order states). The dispersion of lifetimes can be used to infer the origin of broad emission peaks for Si-based samples, so we measured decay times with 10 nm steps in the whole emission range. The rise times are below the temporal resolution of our setup. We show the obtained dependences in Fig. 8, overlaid with PL spectra.

For the reference sample, three distinct regions can be observed. Around the center of the emission, at 0.83 eV, the decay curves can be fitted with a single exponential function, resulting in one decay time of approximately 1.5 ns, related to the radiative lifetime in the QDs. Note that our modeling predicts two distinct lifetimes for the two bright states with different spin configurations (See Tab. III in Sec. IV. C), but they are too close to be resolved in fitting of a typical decay curve. But their average for the ground state is 1.485 ns, which agrees very well with the experimental value. Similar lifetimes have already been reported for excitonic emission in InAs/InP QDs emitting at the third telecom window (see e.g. 1.4-1.5 ns in Ref. 31; 1.59 ns, agreeing with the lifetime calculated with an eight-band *k·p* model, in Ref. 32; or 1.4 ns for 1540 nm emitting InAsP/InP QD in Ref. 33. According to pseudopotential calculations shown in Ref. 34, lifetimes around 1.5 ns indicate dots with heights exceeding 6 nm. For emission energies distant from the PL maximum, we need to use two exponential components to obtain a good fit to experimental data, indicating that a second decay process, with longer times, is present. An analogous situation occurs for all the other registered time evolutions of PL. The second time constant follows the behavior of the shorter one and can be related to the influence of excited or dark exciton states on carrier kinetics, which is naturally more pronounced in larger QDs due to lower energy splittings. However, we will focus only on the interpretation of short decay times, attributed to the characteristic times of radiative recombination processes. In the lower energy region, the measured times decrease slightly with decreasing energy. Such a trend can be explained by the increased average height of the dots, responsible for the energy shift. In the higher energy range, the lifetimes drop considerably, which may be related to the emission of excited states, where nonradiative relaxation to the ground state can dominate the PL decays.

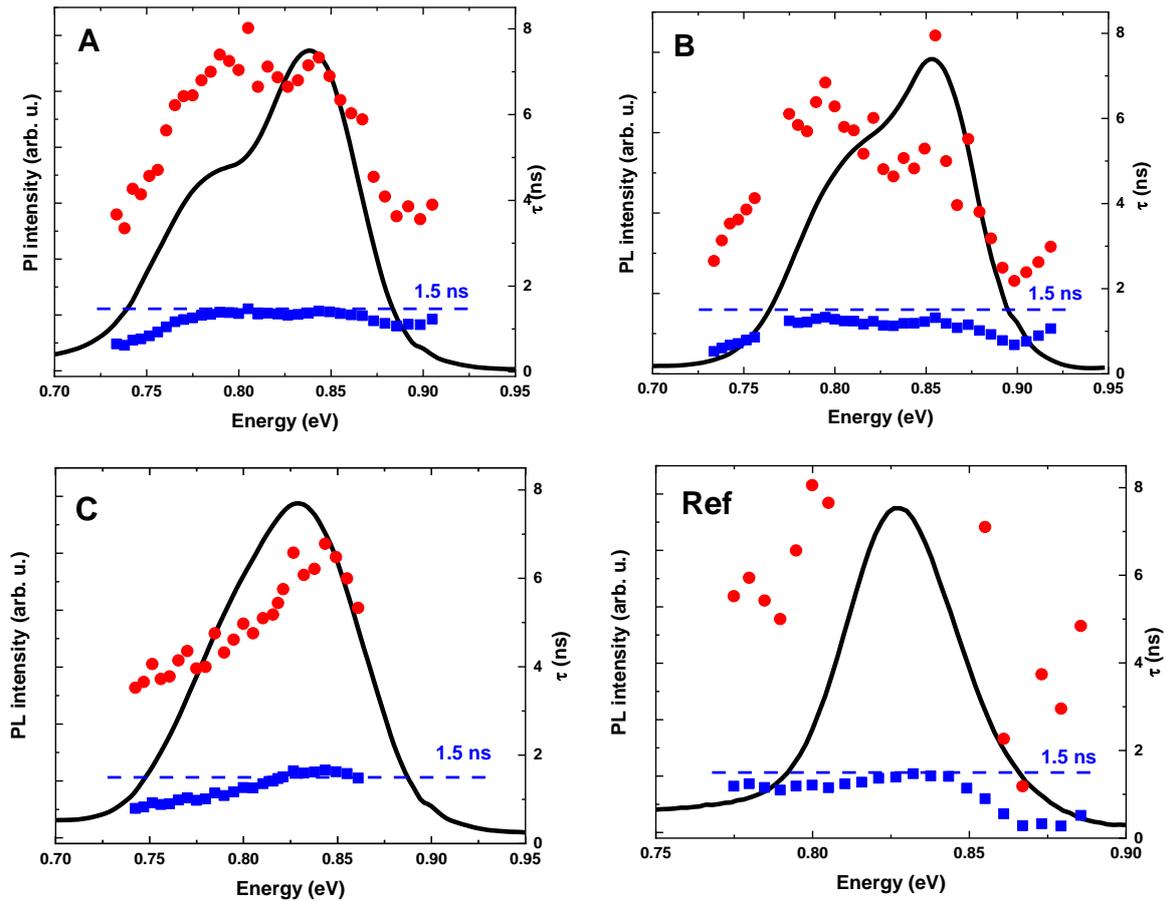

Fig. 8. Dispersion of measured lifetimes superimposed on low-temperature PL spectra for each sample.

The dispersion relations for the Si-based samples are similar in the region around 0.83 eV, in the vicinity of the higher energy PL maximum, with the determined lifetimes of about 1.5 ns. It indicates that this part of the emission spectrum comes from the dots with optical and thus morphological properties similar to those in the InAs/InP reference structure. However, a much more pronounced reduction of lifetimes below 0.8 eV compared to the reference can be related to a population of dots with considerably different geometrical parameters, overlapping with the second PL maximum attributed to other family of QDs. Such a population, especially with optical transitions at lower energy, is unfavorable in the case of dots purposed as an active material for laser devices, constituting additional radiative or non-radiative escape channels for carriers. Therefore, the growth of QDs on silicon should be further optimized, possibly with additional steps aimed at enhanced QD uniformity, to remove those lower-energy emitting nanostructures.

### F. Polarization of emission

As a final characterization tool, we utilize linear-polarization-resolved PL measurements for all the samples. In Fig. 9, we first compare the obtained ellipses of polarization between the reference and the sample A, and then between all the Si-based samples. For the samples A-C, two plots are shown, with PL intensity taken at the maxima of PL peaks. For all the samples the determined angular orientation of the ellipse is, within experimental accuracy, identical. Since the samples are mechanically cleft from wafers before measurements, their edges are aligned along the same crystallographic weak planes (110) and (1-10), enabling direct comparison of polarization angles between the samples. The degree of polarization of emission is affected mostly by the asymmetry of dots, which leads to the mixing between QD-confined heavy and light hole states. All of those factors depend on the confining potential, in turn directly related to the shape and composition of dots, which can be traced back to the growth conditions. The agreement of polarization ellipse angular alignment between the reference and Si-based samples means that during growth the structures retain their crystallographic structure at the transition from diamond to zinc blend structure. It is very promising for potential applications of InAs QDs grown on silicon substrates.

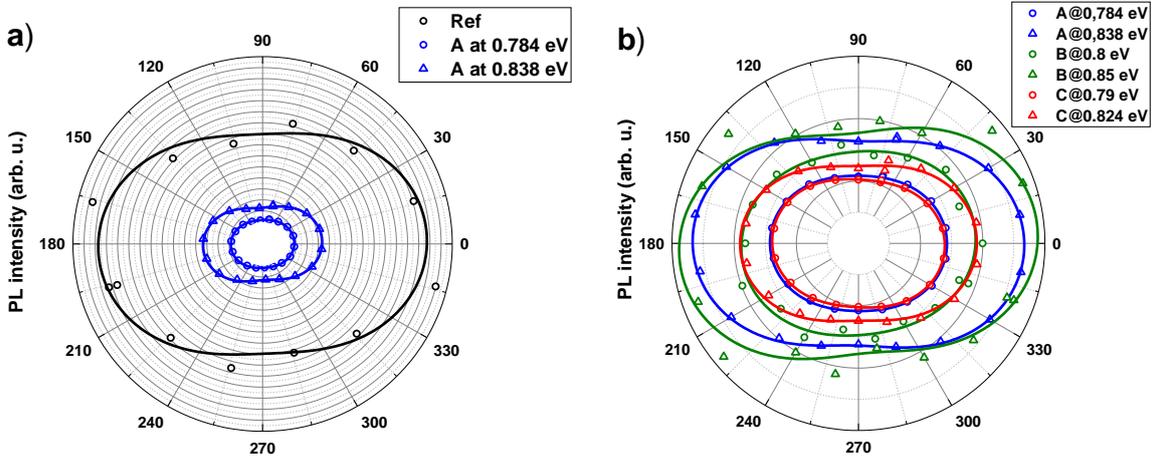

Fig. 9. Linear polarization ellipsis: a) comparison of reference and structure A, b) comparison of all Si-based structures.

To analyze the results quantitatively and to gain more insight into the issue of different QD populations, we calculate the degree of linear polarization (DOP),

$$\text{DOP} = \frac{I_{max}-I_{min}}{I_{max}+I_{min}}, \quad (3)$$

where $I_{max}$ and $I_{min}$ stand for the maximal and minimal PL intensity, respectively. In Fig. 10, we show the dependence of DOP on the emission energy, together with PL intensity.

The DOP for the reference sample is on the level of 0.25 and remains constant for most of the emission peak, only slightly increasing at its high-energy tail, where the lifetimes decrease. In the case of the Si-based structures, two distinct levels of DOP can be seen. In the spectral range of the high-energy peak, for energies close to the reference emission energy, the DOP is above 0.20, which agrees with the value determined in our simulation (see Tab. III in Sec. IV. C). However, for the low-energy peak, DOP drops considerably to the level of 0.10-0.15, suggesting significantly different geometry of these dots, including probable higher in-plane symmetry of confining potential than for the reference sample. It is another suggestion that the Si-based structures indeed host two different populations of dots, and only one of them has properties similar to the QDs grown on InP substrates.

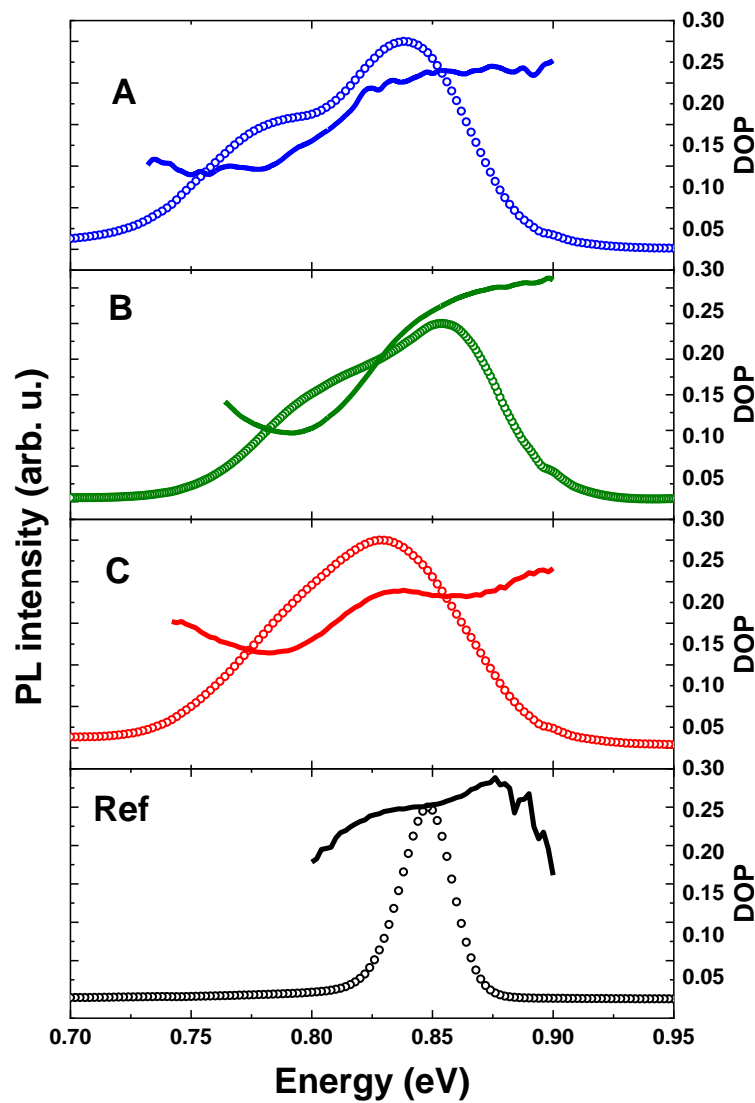

Fig. 10. DOP superimposed on low-temperature PL spectra for all samples.

## V. CONCLUSIONS

We have grown a series of silicon-based InAs QD structures with different defect filtering layers, targeted at telecom wavelength laser applications. We have employed complementary optical characterization techniques to determine their emission properties and confront them with the high-quality reference sample, grown on an InP substrate. Our main positive conclusion is that the used DFLs play their role very efficiently, and the optical properties of Si-based QDs are comparable to the reference QDs and are not affected by defects significantly. There is no considerable difference between DFLs used, however, the careful analysis of optical characterization results summarized in Tab. V shows that samples with top QD DFL exhibit better properties.

Tab. V. Comparison of device-oriented characteristics for Si-based QD samples.

| Sample name | Strain conditions | RT PL | LT PL | Thermal stability | Homogeneity | Total |
|---|---|---|---|---|---|---|
| A | → | ↓ | → | → | ↓ | ③ |
| B | ↑ | → | ↑ | → | ↓ | ② |
| C | ↑ | → | → | ↑ | → | ① |

The sample C, having two QD-based DFLs is the closest to the reference when regarding the main optical and electronic characteristics, and thus most promising for laser applications. Still, there is present a much larger broadening of emission for Si-based structures, due to the presence of an additional lower-energy emission peak, which we attribute to a second family of QDs, with optical properties distinct from the reference structure. Such a population must result from differences in growth conditions related to some residual effect of the silicon substrate on strain which is not fully compensated and leads to deviations from preferable flat surfaces during epitaxial growth. Improvements in the growth procedure are the way to reduce the density of those superfluous dots since they might be considered as a limitation of the applicative potential of the investigated structures.

**NOTES**

The authors have no conflicts to disclose.

**ACKNOWLEDGEMENTS**

The work has been supported by the project no. 2019/33/B/ST5/02941 of OPUS17 call of the National Science Centre in Poland and by European Union Horizon 2020 research and innovation programme under grant agreement No 780537, project MOICANA (Monolithic cointegration of QD-based InP on SiN as a versatile platform for the demonstration of high performance and low cost PIC transmitters, www.moicana.eu). M. G. acknowledges support from the National Science Centre (Poland) under Grant No. 2015/18/E/ST3/00583. M. G. is grateful to Krzysztof Gawarecki for sharing his **k·p** code. Part of the calculations has been carried out using resources provided by the Wroclaw Centre for Networking and Supercomputing (https://wcss.pl), Grant No. 203.